\documentclass[aps,hidelinks,twocolumn, prl, superscriptaddress,float fix]{revtex4-2}
\usepackage{graphicx}
\usepackage{dcolumn}
\usepackage{orcidlink}
\usepackage{caption}
\usepackage{float}
\usepackage{subcaption}
\usepackage{csquotes}
\usepackage{float}
\usepackage{amsfonts}
\usepackage{amsmath}
\usepackage{amsmath,amssymb}
\usepackage[italicdiff]{physics}
\usepackage{bm}
\usepackage{mathtools}
\usepackage{enumitem}
\usepackage{tensor}
\usepackage{wrapfig}
\usepackage{footnote}
\usepackage{here, comment}
\usepackage{bm}
\usepackage{dsfont}

\def\ket#1{\left|#1\right\rangle}
\def\braket#1#2{\left\langle #1|#2\right\rangle}
\def\ketbra#1#2{\left|#1\right\rangle\left\langle#2\right|}
\def\abs#1{\left|#1\right|}

\def\TFD{\text{TFD}}

\begin{document}


\title{Complexity and the Hilbert space dimension of 3D gravity}

\author{Vijay Balasubramanian\,\orcidlink{0000-0002-6497-3819}}
	
\affiliation{David Rittenhouse Laboratory, University of Pennsylvania,  Philadelphia, PA
 19104, USA}

\affiliation{Santa Fe Institute, 1399 Hyde Park Road, Santa Fe, NM 87501, USA}

\affiliation{Theoretische Natuurkunde, Vrije Universiteit Brussel, Pleinlaan 2, B-1050 Brussels, Belgium}

\author{Rathindra Nath Das\,\orcidlink{0000-0002-4766-7705}}
\email{das.rathindranath@uni-wuerzburg.de}

\affiliation{Institute for Theoretical Physics and Astrophysics and W\"urzburg-Dresden Cluster of Excellence ctd.qmat, Julius-Maximilians-Universit\"at W\"urzburg, 
 Am Hubland, 97074 W\"{u}rzburg, Germany}

\affiliation{Faculty of Physics, Weizmann Institute of Science, Rehovot 7610001, Israel}
\affiliation{Center for Theoretical Physics—a Leinweber Institute, MIT,  Cambridge, MA 02139, USA}

\author{Johanna \nolinebreak Erdmenger\,\orcidlink{0000-0003-4776-4326}}

\affiliation{Institute for Theoretical Physics and Astrophysics and W\"urzburg-Dresden Cluster of Excellence ctd.qmat, Julius-Maximilians-Universit\"at W\"urzburg, 
Am Hubland, 97074 W\"{u}rzburg, Germany}

	

\author{Jonathan Karl\,\orcidlink{0009-0000-8729-4213}}
\email{jonathan.karl@uni-wuerzburg.de}
	
\affiliation{Institute for Theoretical Physics and Astrophysics and W\"urzburg-Dresden Cluster of Excellence ctd.qmat, Julius-Maximilians-Universit\"at W\"urzburg, 
 Am Hubland, 97074 W\"{u}rzburg, Germany}

	

\author{Herman Verlinde\,\orcidlink{0009-0003-2023-7179}}

\affiliation{Department of Physics, Princeton University, Princeton, NJ 08544, USA}

\date{\today}

\begin{abstract}
\noindent  A central problem in formulating a theory of quantum gravity is to determine the size and structure of the Hilbert space of  black holes. Here we use a quantum dynamical Krylov complexity approach to calculate the Hilbert space dimension of a black hole in 2+1-dimensional Anti-de Sitter space. We achieve this by obtaining the spread of an initial thermofield double state over the Krylov basis. The associated Lanczos coefficients match those for chaotic motion  on the $SL(2,\mathbb{R})$ group. 
By including non-perturbative effects in the path integral,
which computes coarse-grained ensemble averages, we find that
the complexity saturates at late times. The saturation value is given by the exponential of the Bekenstein-Hawking entropy. Our results introduce a new way to compute the Hilbert space dimension of complex interacting systems from the saturating value of spread complexity.
\end{abstract}

\date{\today}

\maketitle

\footnotetext[1]{See Supplemental Material for a derivation of the Lanczos coefficients following from the MWK density of states \eqref{eq:MWK_density}, a derivation of \eqref{eq:cmkw}, and a derivation of the double pole structure in \eqref{eq:A_double_pole}}

\footnotetext[2]{We denote averaged quantities with an overline $\bar{.}\ $}

\footnotetext[3]{For a theory with a discrete spectrum we have $\rho(E)=\sum_i\delta(E_i-E)$, where the $E_i$ are the energy eigenvalues}

\footnotetext[4]{see \cite{Miyaji:2025yvm,Miyaji:2025jxy} for a definition of the microcanonical TFD state }

\paragraph*{Introduction---} 

An outstanding problem in quantum gravity is to understand the size and structure of  Hilbert spaces  describing different gravitating objects. One way to determine the Hilbert space dimension is to work out the number of vectors in a complete, orthonormal basis.  This is the procedure  used in \cite{STROMINGER199699} to explain the entropy of certain extremal, supersymmetric black holes in string theory.  A second approach is to  consider any sufficiently large collection of distinct vectors.  The rank of the Gram matrix of overlaps of these vectors equals the Hilbert space dimension.  This is the procedure  used in \cite{Balasubramanian:2022gmo,Climent:2024trz,Magan:2024nkr,Banerjee:2024fmh,Balasubramanian:2024rek,Balasubramanian:2025zey, Balasubramanian:2025hns} to account for the  entropy of general black holes. The latter approach employed the Euclidean gravity path integral and its saddlepoint approximation,  assumed to be well defined by some high energy completion, perhaps by a string theory.  There is a third approach which applies to chaotic, or at least sufficiently ergodic, systems:  to compute the dimension of the span of all  states explored by a generic initial vector as it evolves over time.

We apply this third approach to asymptotically AdS$_3$ gravity by computing spread complexity \cite{Balasubramanian:2022tpr} of evol\-ving thermofield double (TFD) states describing near-extremal eternal black holes.  Spread complexity is computed from the support of quantum states on the Krylov basis, constructed to span the  time evolution.  Our Krylov basis will be related to the generalized TFD states in \cite{Magan:2024nkr,Banerjee:2024fmh}.
In chaotic theories, generic initial states are eventually completely deloca\-lized in the Krylov basis \cite{Balasubramanian:2022tpr,Balasubramanian:2023kwd,Erdmenger:2023wjg}, so that measuring the spread reveals the dimension of the accessible part of the Hilbert space. We  work with thermal states of 3d gra\-vity with support on the entire infinite dimensional energy eigenspace, but the exponential suppression of high energies effectively restricts dynamics to a microcanonical band as in statistical mechanics.

We start with the return amplitude, i.e., the probability for a state to remain unchanged, processed into hopping amplitudes, known as  Lanczos coefficients, between a chain of Krylov basis states \cite{Balasubramanian:2022tpr}.  For initial TFD states the return amplitude is  an analytically continued partition sum  determined by the density of states.  Spread complexity can then be written in terms of the spectral density and certain orthogonal polynomials in the energy built from the Lanczos coefficients \cite{Muck:2022xfc,Caputa:2025ozd,Balasubramanian:2025xkj}. Identifying these polynomials as those of a random matrix ensemble, our method introduces a new physical notion of the eigenvalue statistics of the Hamiltonian by means of a random matrix theory (RMT)  with matrix potential $e^{-V(E)} = |\phi_0(E)|^2\rho(E)$ specified by the reference state $\phi_0(E)$ and microscopic spectral density $\rho(E)$.

We apply this procedure to  3d gravity in a near-extremal limit, where the spectral density $\rho(E)$ is known \cite{Maloney:2007ud,Keller:2014xba,Maxfield:2020ale}. This goes beyond the well-understood example of JT gravity in 1+1 dimensions \cite{Iliesiu:2021ari,Balasubramanian:2024lqk,Miyaji:2025yvm}. We show that the Lanczos coefficients  match those of  quantum particles moving on the $SL(2,\mathbb{R})$  manifold, with the associated Laguerre polynomials  leading to  monotonically increasing spread complexity. The spectral density $\rho(E)$ in \cite{Maloney:2007ud,Keller:2014xba,Maxfield:2020ale} was derived from the Euclidean gravity path integral (GPI); many studies, e.g., \cite{Saad:2019lba,Marolf:2020xie,Balasubramanian:2022gmo} and \cite{Cotler:2020ugk} for 3d gravity, suggest that this $\rho(E)$ should be interpreted as a coarse-grained average over an underlying ensemble. As such, products like $\rho(E)\rho(E')$  in the formula for spread complexity  should be understood as spectral correlators in this ensemble rather than as simple products.  We assume that these correlators follow random matrix universality, as suggested by effects of wormholes contributing to the product of partition sums in 2d and  3d gravity \cite{Maxfield:2020ale,Cotler:2020ugk}. The  same wormholes imply a finite-dimensional black hole Hilbert space spanned by time-shifted TFD states \cite{Magan:2024nkr,Banerjee:2024fmh}. Including this effect,   spread complexity saturates at a value exponential in the Bekenstein-Hawking entropy, i.e., the black hole Hilbert space dimension.

\medskip

\paragraph*{Orthogonal polynomials and complexity---} 

The idea behind spread complexity is to quantify the spread of a wavefunction in the Hilbert space as it  evolves over time \cite{Balasubramanian:2022tpr, Nandy:2024htc, Baiguera:2025dkc,Rabinovici:2025otw}. Consider an initial state $\ket{\psi_0}$,  evolving  as 
\begin{equation}
    \ket{\psi(t)}=e^{-itH}\ket{\psi_0}\equiv\sum\limits_n\psi_n(t)\ket{K_n}\,,
    \label{eq:time-evolution}
\end{equation}
 where the Krylov basis $\{\ket{K_n}\,\vert\,n\in\mathbb{N} \}$ is obtained recursively from the set $\{H^n\ket{\psi_0}\,\vert\,n\in\mathbb{N} \}$ through an orthogonalization procedure. By construction, the basis vectors satisfy a recursion relation 
\begin{equation}
    H\ket{K_n}=a_n\ket{K_n}+b_n\ket{K_{n-1}}+b_{n+1}\ket{K_{n+1}}\,,
    \label{eq:recursion_krylov}
\end{equation}
where  $a_n$ and $b_n$ are the Lanczos coefficients. Since the $\ket{K_n}$ are orthonormal, we obtain the coefficients $\psi_n(t)$ of the state \eqref{eq:time-evolution} in the Krylov basis as $\psi_n(t)=\braket{K_n}{\psi(t)}$. The spread complexity is then defined as
\begin{equation}
    C_S(t)=\sum_nn\vert \psi_n(t)\vert^2\,.
    \label{eq:spread_complexity}
\end{equation}
We can also expand \eqref{eq:time-evolution} in the energy basis as
\begin{equation}
    \ket{\psi(t)}=\int dE\,\rho(E)\,\phi_0(E)e^{-iEt}\ket{E}\,,
    \label{eq:expansion_energy_basis}
\end{equation}
where $\rho(E)$ is the spectral density and $\phi_0(E)=\braket{E}{\psi_0}$. For a theory with a discrete spectrum $\rho(E)=\sum_i\delta(E_i-E)$, where the $E_i$ are energy eigenvalues; we are working in a thermodynamic limit and approximating the spectral density as continuous. The energy eigenstates are normalized such that
\begin{equation}
    \braket{E}{E^\prime}=\frac{\delta(E-E^\prime)}{\rho(E)}\,,\quad\mathds{1}=\int dE\rho(E)\ketbra{E}{E}\,.
    \label{eq:energy_basis}
\end{equation}
Projecting (\ref{eq:expansion_energy_basis}) onto $|K_n\rangle$ gives
\begin{equation}
    \begin{split}
        \psi_n(t)=\int dE\,\rho(E)\,\phi_0(E)e^{-iEt}c^*_n(E)\,,
    \end{split}
    \label{eq:wavefunction_energy_basis}
\end{equation}
where $c_n(E)\equiv\braket{E}{K_n}$. It follows from \eqref{eq:recursion_krylov} that the $c_n(E)$ also satisfy a recursion relation \eqref{apx:new_recursion} with initial conditions $c_{-1}(E)=0$ and $c_0(E)=\phi_0(E)$ (also see \cite{Alishahiha:2022anw,Alishahiha:2024vbf}). We show in the appendix that the solution to this recursion relation takes the form
\begin{equation}
 c_n(E)=\phi_0(E)\,h_n(E)\,,
    \label{eq:polynomial}
\end{equation}
where $h_n(E)$ is a polynomial of degree $n$ with coefficients determined purely from the Lanczos coefficients \eqref{apx:coefficients_polynomials}. It follows from the orthonormality of the Krylov basis that 
\begin{equation}
    \begin{split}
        \int dE\,\rho(E)\abs{\phi_0(E)}^2\,h_n^*(E)h_m(E)=\delta_{nm}\,,
    \end{split}
    \label{eq:orthogonal_polynomial}
\end{equation}
which we obtain by inserting the identity  \eqref{eq:energy_basis} into the overlap $\braket{K_n}{K_m}$. Therefore, the polynomials $h_n(E)$ are orthogonal with respect to the measure $\rho(E)\abs{\phi_0(E)}^2$.  
Moreover, from \eqref{eq:wavefunction_energy_basis} and \eqref{eq:polynomial} we determine spread complexity \eqref{eq:spread_complexity} in terms of these  polynomials as \cite{Muck:2022xfc,Caputa:2025ozd,Balasubramanian:2025xkj}
\begin{equation}
    \begin{split}
        C_S(t)&=\int dE\,dE^\prime\,e^{-i(E-E^\prime)t}\,\rho(E^\prime)\rho(E)\\
        &\times\abs{\phi_0(E)}^2\abs{\phi_0(E^\prime)}^2\sum_nn\,h_n(E^\prime)h_n^*(E)\,.
    \end{split}
    \label{eq:spread_complexity_polynomials}
\end{equation}

\paragraph*{Early time complexity growth.---} 
Now consider an eternal black hole in AdS$_{d+1}$.  The  Hartle-Hawking  no-boundary Euclidean path integral  constructs a vacuum wavefunction of the thermofield double (TFD) form
\begin{equation}
    \ket{\TFD(t)}=\frac{1}{\sqrt{Z(\beta)}}\int dE\,\rho(E)e^{-\frac{\beta}{2}E}e^{-iEt}\ket{E}\, \,
    \label{eq:TFD}
\end{equation}
where $Z(\beta)$ is the canonical partition function 
\begin{equation}
Z(\beta)=\int\!dE\,\rho(E)\,e^{-\beta E}\,,
\label{eq:canonical_partition_function}
\end{equation}
and we use the shorthand  $\ket{E}\equiv\ket{E}_L\otimes\ket{E}_R$, as in \cite{Miyaji:2025yvm,Miyaji:2025jxy}.  The Lanczos coefficients can be obtained from  moments of the return amplitude \cite{Balasubramanian:2022tpr}
\begin{equation}
    S_\text{TFD}(t):=\braket{\TFD(0)}{\TFD(t)}=\frac{Z(\beta+it)}{Z(\beta)}\,.
    \label{eq:return_TFD}
\end{equation}
In the effective low-energy description, $Z$ is obtained from the gravitational path integral (GPI) over manifolds with boundary  $S^1_\beta\times S^{d-1}$ \cite{PhysRevD.15.2752}, which computes a trace over the gravity Hilbert space \cite{Balasubramanian:2025hns}.

For general spacetime dimension this path integral has to be approximated by the leading saddle point. However, for $d+1=3$ Maloney--Witten--Keller (MWK) evaluated the  sum over all classical solutions to Einstein's equations with torus boundary conditions \cite{Maloney:2007ud,Keller:2014xba}. The torus parameter is $\tau=\frac{\theta+i\beta}{2}$, where $\theta$ is the periodicity of the spatial circle. $\tau$ is the dual variable to the angular momentum $J$ of the black hole. Close to extremality \cite{Ghosh:2019rcj}, i.e.~$\abs{J}\gg\frac{L}{G_N}$ and $0<E-\abs{J}\ll\frac{L}{G_N}$, where $L$ is the AdS radius and $G_N$ the Newton's constant, the leading contribution to the MWK density of states is \cite{Maxfield:2020ale} 
\begin{equation}
\rho_J^{\text{MWK}}(E)
\;\simeq\;
Ae^{S_0}\,\sqrt{E-\abs{J}}
+
\,\frac{B\,(-1)^J e^{S_0/2}}{\sqrt{E-\abs{J}}} \,,
\label{eq:MWK_density}
\end{equation}
 where $S_0\equiv S_0(J)$ is the Bekenstein-Hawking entropy of the BTZ black hole. The return amplitude \eqref{eq:return_TFD} for the MWK density is obtained from \eqref{eq:canonical_partition_function} and given in the appendix as \eqref{apx:smwk}. As shown in the appendix, the resulting Lanczos coefficients \eqref{apx:an_MWK}-\eqref{apx:bn_MWK} match those for a  particle on the $SL(2,\mathbb{R})$ group manifold when the index $n$ is large \cite{Balasubramanian:2022tpr}.

In fact the MWK density of states cannot be used to define an integration measure \eqref{eq:orthogonal_polynomial} for the polynomials $h_n$, since for exponentially small $E-\abs{J}$ and odd spins, \eqref{eq:MWK_density} becomes negative \cite{Benjamin:2019stq}. To make matters worse, this negativity implies that the underlying theory cannot be unitary. One proposal to resolve the negativity of the density of states  includes additional off-shell contributions to the gravitational path integral \cite{Maxfield:2020ale}. This  shifts the ground state energy and the full spectral density in the near-extremal limit of 3d gravity becomes
\begin{equation}
    \rho^\text{MT}_J(E)\simeq A\, e^{S_0}\sqrt{E-E_0(J)}\,,
    \label{eq:Maxfield_Turiaci_density}
\end{equation}
with $E_0(J)-\abs{J}\propto -(-1)^Je^{-\frac{S_0}{2}}$. Again, the energy density determines the return amplitude \eqref{eq:return_TFD} via \eqref{eq:canonical_partition_function}
\begin{equation}
\;
S_{\text{MT}}(t)
=\left(\frac{\beta}{\beta+i t}\right)^{3/2}
\exp\!\left[\,\!-i\,E_0(J)\,t\right]\,.
\;\label{eq:sfull}
\end{equation}
We can derive  Hamiltonian moments in the initial state and  Lanczos coefficients from  the return amplitude  \cite{Balasubramanian:2022tpr}:
\begin{equation}
    a_n = -\frac{4n+3}{2\beta} - E_0(J)\,, \quad b_n = \begin{cases} 
1 & \text{if } n=0\,, \\
\frac{1}{\beta}\sqrt{\frac{n(2n+1)}{2}} & \text{if } n>0\,, 
\end{cases} \label{eq:LCfull}
\end{equation}
which have  the same form as for particles on  
$SL(2,\mathbb{R})$ \cite{Balasubramanian:2022tpr} 
\begin{equation}
a_n=\gamma(h+n)+\xi\,,\quad b_n=\sigma\sqrt{n(2h+n-1)} \, .
\label{eq:sl2rLanczos}
\end{equation}
The general $SL(2,\mathbb{R})$ case was  worked out in \cite{Caputa:2021sib} by directly solving \eqref{eq:recursion_krylov} for a highest weight state of scaling dimension $h$, with an explicit Hamiltonian. 
One realization, the  harmonic oscillator, gives (\ref{eq:sl2rLanczos}) with $h=1/2$  \cite{Balasubramanian:2022tpr}. Here we find that the partition function of near-extremal 3d gravity leads to \eqref{eq:sl2rLanczos} with $h=3/4$.

The coefficients in \eqref{eq:LCfull} match the $SL(2,\mathbb{R})$ form in \eqref{eq:sl2rLanczos} because the near-extremal limit of 3d gravity has a  dimensional reduction to JT gravity with defects \cite{Maxfield:2020ale}. The reduced theory contains a Schwarzian sector with  $SL(2,\mathbb{R})$ symmetry. In the near-extremal limit the energy density of the Schwarzian theory is approximated by $\sqrt{E-E_\star}$, leading to the first term in \eqref{eq:MWK_density}. On the other hand, from  \cite{Balasubramanian:2022tpr} the spread complexity for the Schwarzian theory coincides with the $SL(2,\mathbb{R})$ model at late times or alternatively in the large $C/\beta$ limit which is the regime where the Schwarzian theory describes JT gravity and $C$ is interpreted as the boundary value of the dilaton.

We show in the appendix that for the Lanczos coefficients \eqref{eq:LCfull} the polynomials in \eqref{eq:polynomial} take the form \eqref{apx:A3_prime}
\begin{equation}
   h_n(E)=\ \sqrt{\frac{\Gamma(\tfrac32)n!}{\Gamma(n+\tfrac32)}}
    L_n^{(1/2)}\!\Big(-\beta(E+E_0(J))\Big)\,,
    \label{eq:cn_closed_form}
\end{equation}
where $L_n^{(\alpha)}$ are the generalized Laguerre polynomials. The polynomials $L_n^{(\alpha)}(y)$ are orthogonal with respect to the measure $\mu(y)\propto y^\alpha e^{-y}$. Therefore, \eqref{eq:cn_closed_form} is indeed consistent with \eqref{eq:orthogonal_polynomial}, since for the TFD state \eqref{eq:TFD} we have $\phi_0(E)\propto e^{-\beta E/2}$, and \eqref{eq:Maxfield_Turiaci_density} implies $\alpha=1/2$. The generalized Laguerre polynomials were used to calculate complexity in \cite{Muck:2022xfc}, and for $\alpha=1/2$ this leads to 
\begin{equation}
    C_S^{\text{MT}}(t)=\frac{3t^2}{2\beta^2}\,. \label{eq:cf}
\end{equation}
This agrees with the complexity growth for a particle on $SL(2,\mathbb{R})$ \cite{Balasubramanian:2022tpr}, after comparing \eqref{eq:sl2rLanczos} to \eqref{eq:LCfull} where we observe that $\gamma^2/4-\sigma^2=0$. 

\paragraph*{Complexity saturation from quantum gravity---}
We expect spread complexity to saturate at late times at a value set by the dimension of the underlying Hilbert space. For black holes we are studying this implies that complexity should saturate at $e^{S_0}$. The reason why \eqref{eq:cf} does not saturate is that we obtained it from a continuous energy spectrum \eqref{eq:Maxfield_Turiaci_density} with infinitely many states in any energy band. By contrast the full quantum gravity theory should have a discrete  spectrum if the finite black hole entropy has a sensible microscopic interpretation.  These considerations suggest that the computations above are only valid for time scales shorter than the inverse of the typical energy spacing $t\ll1/\delta E \sim e^{S_0}$, and that to go to later times we have to find a way to resolve the discreteness of the spectrum. 

The continuum spectral density in \eqref{eq:Maxfield_Turiaci_density} was obtained by  using the gravitational path integral (GPI), an effective field theory description of the underlying unltraviolet complete theory.  By construction, the GPI only contains coarse grained information about the  fundamental degrees of freedom, and is believed to compute quantities that act like averages in some underlying ensemble (see, e.g.,  \cite{Saad:2019lba,Marolf:2020xie,Cotler:2020ugk,Belin:2020hea,Sasieta:2022ksu,Balasubramanian:2022gmo,deBoer:2023vsm}). This suggests that the density of states \eqref{eq:Maxfield_Turiaci_density} should be interpreted as an ensemble average that coarse grains a fundamentally discrete spectrum.

From this perspective, we cannot simply substitute \eqref{eq:Maxfield_Turiaci_density} into the integral expression for spread complexity \eqref{eq:spread_complexity_polynomials}. Rather, if we use the GPI to compute the integrand, the $\rho(E')\rho(E)$ factor  should act like a density-density correlator in the effective ensemble. Thus, if we use the GPI to directly calculate the integrand in the complexity integral, we are effectively computing 
\begin{equation}
    \begin{split}
            \overline{C_S(t)}&=\int dE\,dE^\prime\,e^{-i(E-E^\prime)t}\,\overline{\rho(E^\prime)\rho(E)}\\
            &\times\abs{\phi_0(E)}^2\abs{\phi_0(E^\prime)}^2 A(E,E^\prime)\,,
    \end{split}
    \label{eq:spread_complexity_new}
\end{equation}
where the overbar indicates that the quantity is computed by the GPI, and so acts effectively as an ensemble average.  We also introduced the complexity kernel
\begin{equation}
    A(E,E^\prime)=\sum_nn\,h_n(E^\prime)h_n^*(E)\,.
    \label{eq:Kernel}
\end{equation}
The fine grained structure of the spectrum is encoded in the density-density correlator $\overline{\rho(E')\rho(E)}$.

This spectral correlator can be computed as an inverse Laplace transform of the two-boundary partition function $\overline{Z(\beta^\prime,\beta)}$, where $\beta,\beta^\prime$ are the periodicities of the two thermal circles. The leading path integral contribution to this partition function is disconnected \cite{Banerjee:2024fmh}, resulting in the factorized form $Z(\beta^\prime)Z(\beta)$, leading to spread complexity of the form \eqref{eq:spread_complexity_polynomials}, which does not saturate. The leading non-perturbative correction to the factorized partition function comes from a Euclidean wormhole that connects the two boundaries. For 3d gravity, this contribution has been calculated in \cite{Cotler:2020ugk}. In the near-extremal limit discussed above \eqref{eq:MWK_density} the wormhole contribution has a universal random matrix form \cite{Boruch:2025ilr}, consistent with the fact that the effective dimensionally reduced description has a known random matrix dual in the  Gaussian Unitary Ensemble (GUE) universality class \cite{Maxfield:2020ale}. 

Thus, the spectral correlator in \eqref{eq:spread_complexity_new} shows random matrix universality for $E\approx E^\prime$, so that \cite{Saad:2019lba} 
\begin{equation}    
    \begin{split}
\overline{\rho(E^\prime)\rho(E)}&\approx\rho(E^\prime)\rho(E)-\frac{\sin^2(\pi\rho(\overline{E})(E^\prime-E))}{\pi^2(E^\prime-E)^2}+\dots\,.
    \end{split}
    \label{eq:conn_rho}
\end{equation}
Here $\overline{E}=\frac{E+E^\prime}{2}$ and ``$\dots$'' denotes contributions from  contact terms proportional to $\delta(E-E^\prime)$, which lead to a constant offset in the spread complexity \eqref{eq:spread_complexity_new} that we can drop. The second term in \eqref{eq:conn_rho}, arising from the connected wormhole contribution to the path integral is called the sine kernel, and is universal for any matrix model in the GUE universality class \cite{Saad:2019lba}. Since \eqref{eq:conn_rho} vanishes as  $E\to E^\prime$, the universal form of the spectral correlator indicates level repulsion between the energy eigenvalues, a signature of  spectral discreteness  in the fundamental theory \cite{Miyaji:2025yvm,Miyaji:2025jxy}.

We are now ready to evaluate the complexity integral (\ref{eq:spread_complexity_new}) -- the two key ingredients are the density-density correlator that we just discussed, and the orthogonal polynomials (\ref{eq:cn_closed_form}) from which we can explicitly calculate the complexity kernel (\ref{eq:Kernel}). Note that we are interested in the late time behavior of \eqref{eq:spread_complexity_new}, where the integral is dominated by frequencies $\omega:=E- E^\prime$ of order $1/t$. In this regime the spectral correlator has the form \eqref{eq:conn_rho}, and we show in the appendix that the kernel \eqref{eq:Kernel} is given by 
\begin{equation}
    A(E,E^\prime)\ =\ \frac{\mathcal{C}(\overline{E})}{(E-E^\prime)^2}\,+\,\text{(terms regular in } E-E^\prime)\,,
\label{eq:A_double_pole}
\end{equation}
where $\mathcal{C}$ is a non-zero and smooth function \eqref{eq:C_coeff}.

We show in the appendix that as $E-E^\prime \to 0$ the spectral correlator \eqref{eq:conn_rho} vanishes as $(E-E^\prime)^2$ (see \eqref{apx:expansion_sine}); so the integrand is regular in $\omega = E- E^\prime$ and will thus lead to a finite late time limit of spread complexity. Without the non-perturbative corrections, which result in the sine kernel contribution to the spectral correlator, the double pole in \eqref{eq:A_double_pole} would have led a late time divergence of complexity as in \eqref{eq:cf}. Evaluating the $\omega$ integral in the appendix, following \cite{Iliesiu:2021ari}, where a similar integral appeared in the context of the wormhole length in JT gravity, we find \eqref{apx:energy_integral_lowerlimit_app_rewrite}
\begin{eqnarray}
       \overline{C_S(\hat{t})}& =& C_0+\frac{2\pi^2 e^{S_0}}{3\hat Z(\beta)^2}\int_{\overline{E}_\star(\hat{t})}^\infty d\overline{E}\;
    e^{-2\beta \overline{E}}\;
    \hat\rho(\overline{E})^3\;
    \mathcal{C}(\overline{E}) \nonumber \\
    & &\qquad\qquad  \times\Bigl(1-X(\hat t;\overline{E})\Bigr)^3\,
    \label{eq:complexity_late_time}
\end{eqnarray}
where $e^{S_0}\hat{\rho}(\overline{E})=\rho(\overline{E})$, $e^{S_0}\hat Z(\beta)=Z(\beta)$, and $e^{S_0}\hat{t}=t$. Moreover, $X(\hat t;\bar E)$ and $\overline{E}_\star(\hat{t})$ are defined in \eqref{apx:X_def_app_rewrite} and \eqref{apx:Estar_def_app_rewrite} and $C_0$ is a constant. We show in the appendix that for $\hat{t}\to\infty$ the integral in \eqref{eq:complexity_late_time} vanishes. Therefore, we obtain the late-time saturation value of complexity from the constant term in \eqref{eq:complexity_late_time}, which we determine in the appendix and find \eqref{eq:C0_closed_app_rewrite}
\begin{equation}
   C_0=e^{S_0}\,
    \frac{A\pi^{3}}{8}\,
    (2E_0(J))^{\frac{3}{2}}e^{4\beta E_0(J)}\Gamma\!\Bigl(-\frac32,\,4\beta E_0(J)\Bigr)\,.
    \label{eq:dimension_Hilbert space}
\end{equation}
This saturation relies on random matrix universality \eqref{eq:conn_rho} and the double pole in the complexity kernel \eqref{eq:A_double_pole}, similarly to observations in \cite{Alishahiha:2022anw,Miyaji:2025jxy} motivated by JT gra\-vity and its relation with the double-scaled SYK model
\cite{Iliesiu:2021ari,Miyaji:2025yvm,Miyaji:2025ucp}.
Our final result \eqref{eq:dimension_Hilbert space} shows that the Hilbert space of 3d gravity near extremality has a finite dimension that is exponential in the black hole entropy $S_0$. 

\medskip

\paragraph*{Summary and outlook---} 
We  used spread complexity to study the structure of the black hole Hilbert space in 3d gravity. For generic initial states of chaotic systems with  discrete, bounded spectra, the saturating value of spread complexity is related to the Hilbert space dimension. In the context of gravity, the challenge is that without a complete microscopic theory, we can only compute complexity from the density of states in an effective theory described by the gravitational path integral. As this approach results in a continuous and unbounded spectrum, we find that naively complexity grows for all times, suggesting an infinite dimensional Hilbert space. We overcome this obstacle by extracting information about the discreteness of the spectrum in the microscopic theo\-ry from the density-density correlator appearing in the spectral representation of complexity.  This correlator can be computed from the effective path integral, which ave\-rages over the microscopic details of the fundamental theory. The outcome is saturation of spread complexity at a value exponential in the black hole entropy.

There is a conjecture that in holographic theories of gravity, the complexity of evolving TFD states is geometrized, possibly as the volume of distinguished spacelike slices passing through the Einstein-Rosen bridge in the gravitational description \cite{Susskind:2014rva,Stanford:2014jda}. Alternative proposals suggest that complexity might be related to other diffeomeorphism invariant quantities \cite{Brown:2015bva,Belin:2021bga,Erdmenger:2022lov,Das:2024zuu}.  In most of these proposals, the complexity in question was conceived of in terms of a dual field theory. But since, in holography, the field theory state and the quantum gravity state must be dual to each other, we might as well compute  complexity directly in the bulk gravity if we are able to do it.  The question is what notion of complexity to use.  Here we looked at a measure of complexity associated to the spreading quantum state \cite{Balasubramanian:2022tpr} that has found numerous  applications. In fact,  recently the authors of \cite{Lin:2022rbf,Rabinovici:2023yex} showed that  this notion of spread complexity applied to TFD states of the Double-Scaled SYK model \cite{Berkooz:2018qkz,Berkooz:2018jqr} precisely reproduces the growing classical length of wormholes in the dual 2d Jackiw-Teitelboim gravity.  Then by invoking RMT correlations in the Lanczos coefficients, the authors of  \cite{Balasubramanian:2022dnj,Nandy:2024zcd,Balasubramanian:2024lqk} showed that the spread complexity  saturates at late times, implying large quantum corrections to the classical notion of length.

In a complementary approach, the authors of \cite{Iliesiu:2021ari} defined a non-perturbative notion of  wormhole length directly from the path integral of a  class of 2d dilaton gravity theories, including JT gravity with defects, which is a dimensionally reduced description of the near-extremal sector of 3d gravity \cite{Maxfield:2020ale} (also see  \cite{Miyaji:2025yvm,Miyaji:2025ucp}). The resulting expression for  2d wormhole length is very similar to our result for spread complexity \eqref{eq:complexity_late_time} and only differs in the functional form of the $\mathcal{C}(\overline{E})$ (compare \eqref{eq:C_coeff} and \eqref{eq:complexity_late_time} to Eqs.~3.10 and 4.6 in \cite{Iliesiu:2021ari}).  An explanation for this discrepancy is that the non-perturbative wormhole length in \cite{Iliesiu:2021ari} is defined by summing over geodesics in all Euclidean 2d geometries with a single circular boundary, the simplest example being the disc that analytically continues to the Lorentzian black hole.  If we define the non-perturbative wormhole length in a similar way for 3d gravity, we have to sum over geometries with a single torus boundary $S_\beta^1\times S^1$. Some of these lead to the aforementioned 2d geometries after dimensional reduction. However, there are infinitely many 3d geometries which do not have this 2d interpretation. Therefore, the 2d result of \cite{Iliesiu:2021ari} will receive corrections, which should lead to our 3d formula \eqref{eq:complexity_late_time}. We leave the  non-perturbative study of wormhole length in 3d gravity to future work.

 \begin{acknowledgments}
 \vspace{1em}
 \noindent  \paragraph*{Acknowledgements---} We thank Poulami Nandi for collaboration during early stages of this work, as well as Souvik Banerjee, Ben Craps, Gabriele Di Ubaldo, Felix Haehl, Thomas Kögel, Andrew Sontag and Tom Yildirim for useful discussions. RND, JE and JK are supported by Germany's Excellence Strategy through the Würzburg-Dresden Cluster of Excellence ctd.qmat – Complexity, Topology and Dynamics in Quantum Matter (EXC 2147, project-id 390858490), and by the Deutsche Forschungsgemeinschaft (DFG)  through the Collaborative Research centre \enquote{ToCoTronics}, Project-ID 258499086-SFB 1170. RND is supported by the PRIME programme of the German Academic Exchange Service (DAAD), with funds from the German Federal Ministry of Research, Technology and Space (BMFTR). VB was supported in part by the DOE through DE-SC0013528 and  QuantISED grant DE-SC0020360.  This project was also supported by FWO-Vlaanderen project G0A2226N.
 
 \end{acknowledgments}

\newpage
~
\newpage
\onecolumngrid
\renewcommand{\theequation}{A.\arabic{equation}}
\setcounter{equation}{0}

\section*{Supplemental Material}
\paragraph*{Recursion relations, Lanczos coefficients, and orthogonal polynomials---}

Here we give more details on the recursion relation satisfied by the $c_n(E)$ and show that the general solution is given in terms of orthogonal polynomials \eqref{eq:polynomial}. Afterwards, we show that for the case of 3d gravity, which has the Lanczos coefficients of a particle on $SL(2,\mathbb{R})$ \eqref{eq:sl2rLanczos}, the orthogonal polynomials are given in terms of Laguerre polynomials \eqref{eq:cn_closed_form}. Recall that by definition we have $c_n(E)=\braket{E}{K_n}$ and it follows from \eqref{eq:recursion_krylov} that they satisfy the recursion relation
\begin{equation}
\begin{split}
        E\,c_n(E)=a_n\,c_n(E)+b_n\,c_{n-1}(E)+b_{n+1}c_{n+1}(E)\,,
    \label{apx:new_recursion}
\end{split}
\end{equation}
with initial conditions $c_{-1}(E)=0$ and $c_0(E)=\phi_0(E)$. We now prove inductively that the solution to this recursion relation takes the form \eqref{eq:polynomial}, with a polynomial of degree $n$
\begin{equation}
    h_n(E)=\sum_{k=0}^n\alpha_{nk}E^k\,.
    \label{apx:expansion_polynomial}
\end{equation}
\textit{proof:} We first show that $c_1(E)$ takes the form \eqref{eq:polynomial}, with a polynomial $h_1(E)$ of degree one. Indeed, we find from \eqref{apx:new_recursion}
\begin{equation}
    \begin{split}
            c_1(E)&=b_1^{-1}(E\,c_0(E)-a_nc_0(E)-b_{0}c_{-1}(E))=\frac{\phi_0(E)}{b_1}(E-a_0)\,,
    \end{split}
\end{equation}
and consequently $h_1(E)=(E-a_0)/b_1$. By induction, we now prove that if all $c_n(E)$ take the form \eqref{eq:polynomial} with $h_n(E)$ given by \eqref{apx:expansion_polynomial} up to a fixed finite $n\in\mathbb{N}$, then we have $c_{n+1}(E)=\phi_0(E)h_{n+1}(E)$ with a polynomial $h_{n+1}(E)$ given by \eqref{apx:expansion_polynomial}. Indeed, we find from \eqref{apx:new_recursion}
\begin{equation}
    \begin{split}
            c_{n+1}(E)&=b^{-1}_{n+1}(E\,c_n(E)-a_nc_n(E)-b_{n}c_{n-1}(E))=\frac{\phi_0(E)}{b_{n+1}}\left(\sum_{k=0}^n\alpha_{nk}E^{k+1}-a_n\sum_{k=0}^n\alpha_{nk}E^k-b_n\sum_{k=0}^{n-1}\alpha_{n-1,k}E^k\right)\\
            &=\phi_0(E)\sum_{k=0}^{n+1}\frac{\alpha_{n,k-1}-(a_n\alpha_{nk}+b_n\alpha_{n-1,k})}{b_{n+1}}E^k
            =\phi_0(E)\sum_{k=0}^{n+1}\alpha_{n+1,k}E^k=\phi_0(E)h_{n+1}(E)\,,
    \end{split}
\end{equation}
where we impose $\alpha_{nk}=0$ for $k<0$ and $k>n$. Thus, $c_{n+1}(E)$ takes the form \eqref{eq:polynomial} and $h_{n+1}(E)$ is a polynomial of degree $n+1$, with coefficients satisfying a recursion relation
\begin{equation}
    \alpha_{n+1,k}=\frac{\alpha_{n,k-1}-(a_n\alpha_{nk}+b_n\alpha_{n-1,k})}{b_{n+1}}\,,
    \label{apx:coefficients_polynomials}
\end{equation}
with initial conditions $\alpha_{00}=1$, which concludes the proof. 
\begin{flushright}
    $\square$
\end{flushright}

Now consider the case of 3d gravity. We show that the solutions to \eqref{apx:new_recursion} in this case are given by the generalized Laguerre polynomials \eqref{eq:cn_closed_form}. Hence, we first define the dimensionless shifted energy $x:=\beta(E+E_0(J))$ and insert the explicit form of the coefficients \eqref{eq:LCfull} into \eqref{apx:new_recursion}
\begin{equation}
    \begin{split}
    \sqrt{(n+1)(2h+n)} c_{n+1}(x) = \bigl[x+2(h+n)\bigr]c_n(x)-\sqrt{n(2h+n-1)} c_{n-1}(x)\,, 
    \end{split}
    \label{apx:A1}
\end{equation}
where for us $h=\tfrac34$. To remove the square roots, we define an $n$ dependent prefactor $A_n$ via
\begin{equation}
    \frac{A_{n+1}}{A_n}=\frac{1}{\sqrt{(n+1)(2h+n)}}\,, \label{apx:A_n_recursion}
\end{equation}
which is solved by
\begin{equation}
    A_n=A_0\sqrt{\frac{\Gamma(2h)}{n!\Gamma(n+2h)}}\,,
    \label{apx:A_n_def}
\end{equation}
where we used $\prod_{j=0}^{n-1}(j+1)=n!$ and $\prod_{j=0}^{n-1}(2h+j)=\Gamma(n+2h)/\Gamma(2h)$. We set $c_n(x)=A_n P_n(x)$ and \eqref{apx:A1} becomes
\begin{equation}
    P_{n+1}(x)=\bigl[x+2(h+n)\bigr]P_n(x)-n(n+2h-1)P_{n-1}(x)\,,
    \label{apx:A2}
\end{equation}
which we compare with the standard recursion relation satisfied by the generalized Laguerre polynomials
\begin{equation}
    (n+1)L^{(\alpha)}_{n+1}(y)=(2n+1+\alpha-y)L^{(\alpha)}_n(y) - (n+\alpha)L^{(\alpha)}_{n-1}(y)\,.
    \label{apx:recursion_Laguerre}
\end{equation}
This implies
\begin{equation}
    P_n(x)= n! L_n^{(\alpha)}(-x)\,,
\end{equation}
with $\alpha:=2h-1$.
Thus, we find the closed form of the solution to \eqref{apx:new_recursion} from \eqref{apx:A_n_def} as
\begin{equation}
    c_n(E) = A_0
    \sqrt{\frac{\Gamma(2h)n!}{\Gamma(n+2h)}}
    L_n^{(2h-1)}\!\Bigl(-\beta(E+E_0(J))\Bigr)\;, \label{apx:A3}
\end{equation}
where the prefactor $A_0$ is fixed through the initial condition $A_0=c_0(E)=\phi_0(E)$. Moreover, for our case we have $h=\tfrac34$, implying that
\begin{equation}
    c_n(E) = \phi_0(E)
    \sqrt{\frac{\Gamma(\tfrac32)n!}{\Gamma(n+\tfrac32)}}
    L_n^{(1/2)}\!\Bigl(-\beta(E+E_0(J))\Bigr)\;, \label{apx:A3_prime}
\end{equation}
which solves \eqref{apx:new_recursion} for arbitrary initial $\phi_0(E)$. \\

\paragraph*{Lanczos coefficients of the MWK density---}

Here we give more details about the Lanczos coefficients which we obtain from the MWK density of states \eqref{eq:MWK_density}. First, we determine the MWK partition function from \eqref{eq:canonical_partition_function} and find
\begin{equation}
\;
Z_{\text{MWK}}(\beta)=e^{-\beta \abs{J}}\left[\frac{A\,e^{S_0}\Gamma\bigr(\frac{3}{2}\bigl)}{\beta^{3/2}}+\frac{B\,(-1)^Je^{S_0/2}\Gamma(\frac{1}{2}\bigl)}{\beta^{1/2}}\right]\,.
\;
\end{equation}
Thus, we obtain the return amplitude from \eqref{eq:return_TFD} as
\begin{equation}
S_{\text{MWK}}(t)
=\frac{e^{-i\abs{J} t}}{1+\delta}\left[\Big(\dfrac{\beta}{\beta+it}\Big)^{3/2}
+\delta\cdot\Big(\dfrac{\beta}{\beta+it}\Big)^{1/2}\right]\,,
\label{apx:smwk}
\end{equation}
where we set $\delta=2\frac{B}{A}(-1)^Je^{-S_0/2}\beta$. From this return amplitude we determine the Lanczos coefficients
\begin{equation}
    \begin{gathered}
        a_0=-\abs{J}-\frac{1}{2\beta}\left(1+\frac{2}{1+\delta}\right)\,,\\
a_n=-\abs{J}-\frac{1}{2\beta}\Biggl[(4n+1)
-(2n-1)(2n)\frac{Q_{n-2}(\delta)}{Q_{n-1}(\delta)}+(2n+1)(2n+2)\frac{Q_{n-1}(\delta)}{Q_{n}(\delta)}\Biggr]\,,\quad n\ge1\,,
    \end{gathered}
    \label{apx:an_MWK}
\end{equation}
and 
\begin{equation}
b_0=1\,, \qquad
b_n=\frac{1}{\beta}\,
\sqrt{\frac{n(2n-1)}{2}\;\frac{Q_{n-2}(\delta)\,Q_{n}(\delta)}{Q_{n-1}(\delta)^{2}}}\,,
\quad n\ge1\,,
\label{apx:bn_MWK}
\end{equation}
where we introduce the functions
\begin{equation}
    \begin{split}
        Q_n(\delta) :&= 2^{n+1}(n+1)!\,L_{n+1}^{(-1/2)}\left(-\frac{\delta}{2}\right)\,,
    \end{split}
\end{equation}
for $n\ge-1$. We now show that these Lanczos coefficients approach \eqref{eq:sl2rLanczos} for large $n$. Therefore, we consider the asymptotic behavior of the ratio 
\begin{equation}
    R_n:=\frac{Q_{n-1}(\delta)}{Q_n(\delta)}=\frac{1}{2n}\left(1-\frac{\sqrt{\delta/2}}{\sqrt{n}}+\frac{\delta-1}{2n}+O(n^{-3/2})\right)\,,
\end{equation}
where we use the standard asymptotics of the Laguerre polynomials. We thus find the large $n$ behavior of the Lanczos coefficients as
\begin{equation}
a_n = -\frac{4n+3}{2\beta}-\abs{J}\,,\quad b_n = \frac{n}{\beta}\,.
\end{equation}
This agrees with the form of the Lanczos coefficients for a particle on the $SL(2,\mathbb{R})$ group manifold  \cite{Balasubramanian:2022tpr},
 \begin{equation}
    a_n=\gamma(h+n)+\xi,\qquad b_n=\sigma\sqrt{n(2h+n-1)}\;,
\end{equation}
evaluated at large $n$. \\

\paragraph*{Calculation of the spread complexity kernel---}

In this appendix we give a detailed derivation of the complexity kernel \eqref{eq:Kernel} and show that it has a double pole in $\omega=E-E^\prime$. Therefore, we introduce the generating function 
\begin{equation}
    F(s;E,E^\prime):=\sum_{n=0}^{\infty} e^{-s n}h_n(E)h_n(E^\prime)\,, \label{apx:F_def}
\end{equation}
which is related to \eqref{eq:Kernel} by
\begin{equation}
     A(E,E^\prime)=-\partial_s F(s;E,E^\prime)\Big|_{s=0^+}\,. \label{apx:A_from_F}
\end{equation} 
Note that the orthogonal polynomials in \eqref{apx:F_def} are given by generalized Laguerre polynomials \eqref{eq:cn_closed_form}. Thus, we use the Hille-Hardy kernel to evaluate the generating function \eqref{apx:F_def} explicitly as
\begin{equation}
    \begin{split}
            &F(s;x,y)=\Gamma\!\big(\tfrac32\big)\sum_{n=0}^{\infty}\frac{n!\cdot e^{-sn}}{\Gamma\!\big(n+\tfrac32\big)}L_n^{(1/2)}(-x)L_n^{(1/2)}(-y)
    =\Gamma\!\big(\tfrac32\big)\frac{\exp\left(\frac{x+y}{e^s-1}\right)}{\big(1-e^{-s}\big)\big(xy\,e^{-s}\big)^{1/4}}\,
    I_{1/2}\!\!\left(\frac{2\sqrt{xy\,e^{-s}}}{1-e^{-s}}\right)\,, 
    \end{split}
\label{apx:F_explicit}
\end{equation}
where we again set $x:=\beta(E+E_0(J))$, and $y:=\beta(E^\prime+E_0(J))$. We set $t=e^{-s}$ and evaluate the right hand side of \eqref{apx:A_from_F} through
\begin{equation}
    -\partial_s\log(F)= t\partial_t\log F =
    \frac{t}{1-t}
    -\frac{t(x+y)}{(1-t)^2}
    -\frac{1}{4}
    +\,\sqrt{xy}\,\frac{t^2+t}{(t-1)^2\sqrt{t}}\, \frac{I'_{1/2}\left(\frac{2\sqrt{xy\,t}}{1-t}\right)}{I_{1/2}\left(\frac{2\sqrt{xy\,t}}{1-t}\right)}\,,
\end{equation}
where we set $I^\prime_{1/2}(u)=\frac{d}{du}I_{1/2}(u)$.  We don’t actually need the full closed form of those algebraic pieces; what we need is the singular behavior as $x\to y$. Therefore, we set $x=z+\frac{\beta\omega}{2}$, and $y=z-\frac{\beta\omega}{2}$ with $\omega:=E-E^\prime$ and $z=\frac{x+y}{2}$. Then, as $t\to 1^-$, i.e., $s\to 0^+$,
\begin{equation}
    \begin{split}
           \frac{2\sqrt{xyt}}{1-t}
    =\frac{2z}{1-t}\sqrt{1-\frac{\beta^2\omega^2}{4z^2}+O(\omega^3)}\sim\ \frac{2z}{1-t}-\frac{\beta^2\omega^2}{4(1-t)z}+O(\omega^3).
    \end{split}
\end{equation}
Using the closed form of the Bessel function $I_{1/2}(u)=\sqrt{\frac{2}{\pi u}}\sinh u$ and $\sinh u\sim \tfrac12 e^{u}$ we find the characteristic edge exponent 
\begin{equation}
    \begin{split}
            e^{\frac{t(x+y)}{1-t}}\sinh \left(\frac{2\sqrt{xy\,t}}{1-t}\right)&\sim\,\tfrac12\exp\!\left[\frac{\beta^2\omega^2}{4z(1-t)}\right].
    \end{split}
\end{equation}
Hence, up to an overall smooth prefactor $G(z)$ that stays finite and nonzero as $t\to 1$ and $\omega\to 0$, we have
\begin{equation}
    F(t;z,\omega)\ \sim\ \frac{G(z)}{\sqrt{1-t}}
    \exp\!\left[-\frac{\beta^2\omega^2}{4z(1-t)}\right],\quad\text{with}\quad
    (1-t)=1-e^{-s}\sim s.
\end{equation}
 From \eqref{apx:A_from_F} we find the leading singular term of the complexity kernel \eqref{eq:Kernel}, as
\begin{equation}
    A(z,\omega)\ \sim\ \frac{G(z)}{(1-t)^{3/2}}
    \left[1+\frac{\beta^2\omega^2}{2z(1-t)}+\cdots\right]
    \exp\!\left[\frac{\beta^2\omega^2}{4z(1-t)}\right].
\end{equation}
In the strict $s\to 0^+$ limit, with fixed $\omega$ we find that the kernel has a double pole in $\omega$
\begin{equation}
    A(\overline{E},\omega)\ =\ \frac{\mathcal{C}(\overline{E})}{\omega^2}\ +\ \text{(terms regular in }\omega)\,,  \label{eq:dp}
\end{equation}
with a non-zero, smooth coefficient 
\begin{equation}
    \mathcal{C}(\overline{E})=-\frac{\pi}{8}\ \frac{\Gamma(\tfrac32)}{\beta^2z}\ =- \frac{\pi}{8}\ \frac{\sqrt{\pi}/2}{\beta^3(\overline{E}+E_0(J))}\ , \label{eq:C_coeff}
\end{equation}
where $z=\beta(\overline{E}+E_0(J))$. \\

\paragraph{Late-time behavior of spread complexity---}

Here we analyze the late-time behavior of \eqref{eq:spread_complexity_new} and show that it saturates at a finite number, which is exponential in the black hole entropy. From \eqref{eq:TFD} and \eqref{eq:A_double_pole} we find that the spectral representation of spread complexity is
\begin{equation}
                \overline{C_S(t)}=\frac{1}{Z(\beta)^2}\int dE\,dE^\prime\,e^{-i(E-E^\prime)t}\,e^{-\beta(E+E^\prime)}\,\overline{\rho(E^\prime)\rho(E)}\, \frac{\mathcal{C}(\overline{E})}{(E-E^\prime)^2}\,,
    \label{apx:complexity_spectral_repr}
\end{equation}
where $\mathcal{C}$, which only depends on $\overline{E}=(E+E^\prime)/2$, is given by \eqref{eq:C_coeff}. Due to the oscillatory factor $e^{-i(E-E^\prime)t}$ the late time behavior of the integral is dominated by small energy differences $\omega=E-E^\prime\sim1/t$. In this regime, the spectral correlator has the universal form \eqref{eq:conn_rho}, where for now we keep the density of states arbitrary. Introducing the rescaled variable $\hat{\omega}=\omega t$, \eqref{apx:complexity_spectral_repr} is given by 
\begin{equation}
    \overline{C_S(t)}=C_0+\frac{t}{ Z(\beta)^2}\int d\overline{E}\;
    e^{-2\beta \bar E}\;
    \rho(\bar E)^2\;
    \mathcal{C}(\bar E)\int d\hat{\omega}\,e^{i\hat{\omega}}\frac{1}{\hat{\omega}^2}\left[
        1-
        \left(\frac{t}{\pi\rho(\bar E)\hat\omega}\right)^2\sin^2\left(\frac{\pi\rho(\bar E)\hat\omega}{t}\right)
    \right]\,,
\end{equation}
where the constant term arises from contact terms in the spctral correlator and we fix it below such that $\overline{C_S(0)}=0$. We also used that for small energy differences $\overline{E}\approx E\approx E^\prime$. Although the $\hat{\omega}$ integral contains $1/\hat\omega^2$, the integrand is finite at $\hat\omega=0$, since the expression in brakets has the small $\hat{\omega}$ expansion
\begin{equation}
        1-\left(\frac{t}{\pi\rho(\bar E)\hat\omega}\right)^2\sin^2\left(\frac{\pi\rho(\bar E)\hat\omega}{t}\right)=\frac{1}{3}\left(\frac{\pi\rho(\bar E)\hat\omega}{t}\right)^2+O(\hat{\omega}^4)\,,
        \label{apx:expansion_sine}
\end{equation}
and thus cancels the pole at $\hat{\omega}=0$. The $\hat{\omega}$ integral has been solved in \cite{Iliesiu:2021ari}, where it appeared in the definition of the non-perturbative wormhole length for a large class of 2d dilaton gravity models. Therefore, we obtain 
\begin{equation}
\label{apx:energy_integral_lowerlimit_app_rewrite}
    \overline{C_S(\hat t)}
    =
    C_0
    +\frac{2\pi^2 e^{S_0}}{3\hat{Z}(\beta)^2}
    \int_{\bar E_*(\hat t)}^{\infty} d\bar E\;
    e^{-2\beta \bar E}\;
    \hat\rho(\bar E)^3\;
    \mathcal{C}(\bar E)\;
    \Bigl(1-X(\hat t;\bar E)\Bigr)^3\,.
\end{equation}
where we introduce the rescaled quantities $\rho(\bar E)=e^{S_0}\hat{\rho}(\bar E)$, $Z(\beta)=e^{S_0}\hat{Z}(\beta)$, and $t=e^{S_0}\hat{t}$ to show the dependence on the black hole entropy $S_0$ explicitly. We also have
\begin{equation}
\label{apx:X_def_app_rewrite}
    X(\hat t;\bar E)=\frac{\hat t}{2\pi\hat\rho(\bar E)}\,,
\end{equation}
and $\bar E_*(\hat t)$ is defined implicitly by
\begin{equation}
\label{apx:Estar_def_app_rewrite}
    \pi\hat\rho(\bar E_*)=\hat t\,.
\end{equation}
For the Maxfield-Turiaci density \eqref{eq:Maxfield_Turiaci_density} we have $\bar E_*(\hat t)=E_0(J)+\Bigl(\frac{\hat t}{\pi A}\Bigr)^2\,$.
Thus, at late times $\hat t\to\infty$, the lower limit $\bar E_*(\hat t)$ tends to infinity, and the integral in \eqref{apx:energy_integral_lowerlimit_app_rewrite} vanishes. Therefore, we determine the late-time saturation value of \eqref{apx:energy_integral_lowerlimit_app_rewrite} from the constant term, which has to be fixed such that $\overline{C_S(0)}=0$. For $\hat t\to 0$ we have $\bar E_*(\hat t)\to E_0(J)$ as well as
$X(\hat t;\bar E)\to 0$, and we find that the constant is given by
\begin{equation}
\label{eq:C0_def_app_rewrite}
    C_0
    =
    -\frac{2\pi^2 e^{S_0}}{3\,\hat Z(\beta)^2}
    \int_{E_0(J)}^{\infty} d\bar E\;
    e^{-2\beta\bar E}\;
    \hat\rho(\bar E)^3\;
    \mathcal{C}(\bar E)\,.
\end{equation}
Using \eqref{eq:C_coeff} and the Maxfield-Turiaci density of states \eqref{eq:Maxfield_Turiaci_density} a straightforward evaluation of the integral in \eqref{eq:C0_def_app_rewrite} gives
\begin{equation}
\label{eq:C0_closed_app_rewrite}
    C_0
    =
    e^{S_0}\,A\,
    \frac{\pi^{3}}{8}\,
    (2E_0(J))^{3/2}\,
    e^{4\beta E_0(J)}\,
    \Gamma\!\left(-\frac32,\,4\beta E_0(J)\right)\,.
\end{equation}
For $E_0(J)\beta\ll 1$, this further simplifies to
\begin{equation}
\label{eq:C0_smallE0_final_app_rewrite}
    C_0 \simeq
    e^{S_0}\,A\,
    \frac{\pi^3}{12}\,
    (2\beta)^{-3/2}\,.
\end{equation}

\bibliography{biblio}
\end{document}